\documentclass{ws-mpla}
\usepackage[super,compress]{cite}
\usepackage{graphicx}

\newcommand{\ib}{\mathrm{i}}
\newcommand{\kb}{\bm{k}}
\newcommand{\Rb}{\mathbf{R}}
\newcommand{\Tb}{\mathbf{T}}
\newcommand{\hb}{\bm{h}}

\newcommand{\eb}{\bm{e}}
\newcommand{\vb}{\bm{v}}
\newcommand{\vbh}{\hat{\bm{v}}}
\newcommand{\adb}{\allowdisplaybreaks } 
\newcommand{\ann}{\adb \nonumber \\}

\begin{document}

\markboth{N. Emelianova, I. V. Fialkovsky \& N. Khusnutdinov}{Casimir effect for biaxial anisotropic plates with surface conductivity}

\catchline{}{}{}{}{}

\title{CASIMIR EFFECT FOR BIAXIAL ANISOTROPIC PLATES WITH SURFACE CONDUCTIVITY}
\author{N. EMELIANOVA}
\address{Centro de Matem\'atica, Computa\c{c}\~ao e Cogni\c{c}\~ao, UFABC, 09210-170 Santo Andr\'e, SP, Brazil \\
	natalia.emelianova@ufabc.edu.br}
\author{I. V. FIALKOVSKY}
\address{Physics Department, Ariel University, Ariel 40700, Israel
\footnote{On leave of absence from Centro de Matem\'atica, Computa\c{c}\~ao e Cogni\c{c}\~ao, UFABC, 09210-170 Santo Andr\'e, SP, Brazil, ifialk@gmail.com}}
\author{N. KHUSNUTDINOV}
\address{Centro de Matem\'atica, Computa\c{c}\~ao e Cogni\c{c}\~ao, UFABC, 09210-170 Santo Andr\'e, SP, Brazil \\ and Institute of Physics, Kazan Federal University, Kremlevskaya 18, Kazan, 420008, Russia\\ nail.khusnutdinov@gmail.com}

\maketitle

\pub{Received (Day Month Year)}{Revised (Day Month Year)}

\begin{abstract}
The Casimir energy is constructed for a system consisting of two semi-infinite slabs of anisotropic material.  Each of them is characterized by bulk complex dielectric permittivity tensor and surface conductivity on the free boundary. We found general form of the scattering matrix and Fresnel coefficients for each part of the system by solving Maxwell equations in the anisotropic media. 
\keywords{Casimir energy; biaxial anisotropy.}
\end{abstract}  

\ccode{03.70.+k, 03.50.De}

\section{Introduction}

Last decade, the great interest was connected with $2D$ systems due to discovering graphene \cite{Novoselov:2005:Tac} (see, for example last reviews \cite{Woods:2016:MpoCavdWi,Khusnutdinov:2019:cei2dmmr}).  In the same time, many interesting and non-trivial $3D$ materials appear like metamaterials, three-dimensional topological and Chern insulators, Dirac and Weyl semi-metals, all highly anisotropic. 

Different aspects of Casimir effect involving these anisotropic media of different complexity were previously studied. In particular, interaction of passive uniaxial and biaxial media with one of the optical axes being perpendicular to the interface, \cite{Barash1973,Rosa2008,Antezza2017} anisotropic single-negative metamaterials \cite{Zeng2013} 
were among the subjects of research, to mention just a few. However, bianisotropic optically active media with arbitrary orientation of the axes has not yet been considered, while such materials (e.g. TaAs, Na$_3$Sb, MoTe$_2$, etc.) are now under active research\cite{Armitage}.
In this paper we consider the Casimir energy for two anisotropic objects with planar symmetry and surface conductivity. The formulas obtained may be applied for the above noted $3D$ materials.

Using a scattering matrix approach, \cite{Lambrecht:2008:ciodg,Fialkovsky:2018:qfcrbcs} the Casimir energy can be given as 
\begin{equation}\label{eq:R}
	\mathcal{E}_{\textsf{\scriptsize C}} =\iint \frac{d^2 \kb_\perp}{2(2\pi)^3}\int_{-\infty}^{\infty}d\xi\ln \det \left(1 - e^{-2d\kappa}\mathbf{R}'_{1} \mathbf{R}_{2}\right), 
\end{equation}
where, $\kappa\equiv \sqrt{\xi^2 + \kb_\perp^2}$, $k_3=\ib \kappa\, \mathrm{sign}\,\xi$, and $\mathbf{R}'_1(\ib\xi, \kb_\perp), \mathbf{R}_2(\ib\xi,\kb_\perp)$ are the Fresnel reflection matrices. Prime means the opposite direction of scattering. \cite{Fialkovsky:2018:qfcrbcs}.

\section{Scattering Problem}
\begin{figure}[t]
	\centering \includegraphics[width=7 truecm]{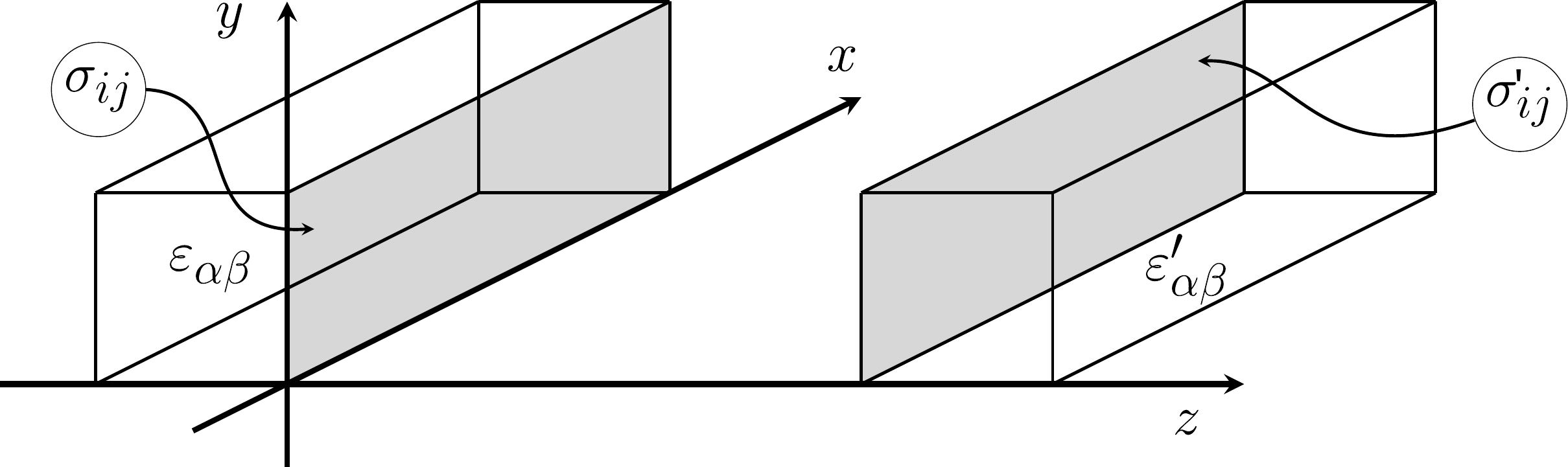}
	\caption{Two dielectric semi-spaces $z<0$ and $z>a$ with dielectric tensors $\varepsilon_{\alpha\beta}$ and $\varepsilon'_{\alpha\beta}$ and~boundaries 
	with surface conductivities $\sigma_{ij}$ and $\sigma'_{ij}$. Magnetic properties we assume to be trivial, 
	$\mu^({}'{}^)=1$.}\label{fig:f1}
\end{figure}

Let us consider the Casimir energy for the system plotted in Fig. \ref{fig:f1}. We consider first a general scattering problem with matter described by hermitian tensor\footnote{The Greek indexes $\alpha,\beta$ run from $1$ to $3$ and the Latin ones run from $1$ to $2$.} $\varepsilon_{\alpha\beta}$ in the left (index $l$) of the boundary $z=0$ and vacuum, $\varepsilon_{\alpha\beta} = \delta_{\alpha\beta}$, in the right (index $r$), which corresponds to the left part of the system in Fig. \ref{fig:f1}. 

Presence of imaginary part of $\bm{\varepsilon}$ (corresponding to optically active media) makes it impossible to find an orthogonal coordinate system where dielectric permittivity tensor would be diagonal. However, it is still perfectly possible to solve Maxwell equation. Generally speaking, Maxwell equations in anisotropic media give a dispersion relation which has $4$  distinctive roots, $\kappa_n = k_3 (\bm{k},\omega)$ and $4$ corresponding distinct eigenvectors $\bm{\mathcal{E}}_{n}$ and $\bm{\mathcal{H}}_{n}$ ($n=1,2,3,4$). We choose numeration of roots such that in the vacuum limit $\kappa_{1,2}\to + k_z$ and $\kappa_{3,4} \to -k_z$ ($k_z = \sqrt{\omega^2 - \bm{k}^2}$). 

The field has the following structure at the left of the boundary (inside matter): 
\begin{equation*}
	(\mathbf{E}_l,\mathbf{H}_l) = e^{\ib \kappa_1 z} A_{i}^{l} (\bm{\mathcal{E}}_{1}^l,\bm{\mathcal{H}}_{1}^l) + e^{\ib \kappa_2 z} B_{i}^{l} (\bm{\mathcal{E}}_{2}^l,\bm{\mathcal{H}}_{2}^l)  + e^{\ib \kappa_3 z} A_{o}^{l} (\bm{\mathcal{E}}_{3}^l,\bm{\mathcal{H}}_{3}^l) + e^{\ib \kappa_4 z} B_{o}^{l} (\bm{\mathcal{E}}_{4}^l,\bm{\mathcal{H}}_{4}^l), 
\end{equation*}
and on its right (in vacuum)
\begin{equation*}
	(\mathbf{E}_r,\mathbf{H}_r) = e^{\ib k_z z} \left\{ A_{o}^{r} (\bm{\mathcal{E}}_{1}^r,\bm{\mathcal{H}}_{1}^r) + B_{o}^{r} (\bm{\mathcal{E}}_{2}^r,\bm{\mathcal{H}}_{2}^r) \right\}  + e^{-\ib k_z z} \left\{ A_{i}^{r} (\bm{\mathcal{E}}_{3}^r,\bm{\mathcal{H}}_{3}^r) + B_{i}^{r} (\bm{\mathcal{E}}_{4}^r,\bm{\mathcal{H}}_{4}^r) \right\}, 
\end{equation*}
where the subscript $i(o)$ denotes incoming (outgoing) wave on the boundary.  We have $8$ amplitudes, $A_{i,o}^{l,r}, B_{i,o}^{l,r}$ to be defined.  They are related by the scattering matrix which is to be defined in its turn through boundary (matching) conditions.

The \textit{in} and \textit{out} states and $\mathbf{S}$-matrix have the following form: $\mathbf{E}^{out} = \left[ A^l_o, B^l_o, A^r_o, B^r_o \right]^{T}$, $\mathbf{E}^{in} = \left[A^l_i, B^l_i, A^r_i, B^r_i \right]^{T}$, $\mathbf{E}^{out}  = \mathbf{S} \cdot \mathbf{E}^{in}$,
where 
\begin{equation}\label{eq:S}
	\mathbf{S} = 
	\begin{bmatrix}
		\Rb & \Tb' \\
		\Tb & \Rb'
	\end{bmatrix},\ \Rb = 
	\begin{bmatrix}
		r_{xx} & r_{xy} \\
		r_{yx} & r_{yy}
	\end{bmatrix}, \ 
	\Tb = 
	\begin{bmatrix}
		t_{xx} & t_{xy} \\
		t_{yx} & t_{yy}
	\end{bmatrix}.
\end{equation}
To obtain $\mathbf{S}$-matrix, we shall use explicit expression for $\mathbf{E}_{r,l},\mathbf{H}_{r,l}$, obtained in the next Section, and impose on them the following boundary conditions:
\begin{equation}\label{eq:boundary}
	\left.(\mathbf{E}^l - \mathbf{E}^r)\times \bm{n}_{l\to r}\right|_{z=0} =\bm{0},\ \left.(\mathbf{H}^l - \mathbf{H}^r)\times \bm{n}_{l\to r} \right|_{z=0} = 4\pi \left.\bm{\sigma}_s \mathbf{E}^r\right|_{z=0}.
\end{equation}
 
\section{Maxwell Equation in Media and the $\mathbf{S}$-matrix}
Let us seek the solutions to the Maxwell equations in the plane waves form $(\mathbf{E,H}) = e^{\ib k_1 x + \ib k_2 y + \ib k_3 z -\ib \omega t} (\bm{\mathcal{E,H}})$, with constant amplitudes $\bm{\mathcal{E}}$ and $\bm{\mathcal{H}}$. 

The equations can be represented in the form of an eigenproblem $\mathbf{M}\cdot \bm{v} = k_3 \bm{v}$, where the matrix $\mathbf{M}$ and   $\bm{v}$ are given by ($e_{\alpha\beta}$ is minor of element $(\alpha,\beta)$ in $\bm{\varepsilon}$)
\begin{equation}\label{eq:M}
	\mathbf{M} = 
	\begin{bmatrix}
		-k_1 \frac{\varepsilon_{31}}{\varepsilon_{33}} & -k_1 \frac{\varepsilon_{32}}{\varepsilon_{33}} & \frac{k_1 k_2}{\omega \varepsilon_{33}} & \omega - \frac{k_1^2}{\omega \varepsilon_{33}} \\
		-k_2 \frac{\varepsilon_{31}}{\varepsilon_{33}} & -k_2 \frac{\varepsilon_{32}}{\varepsilon_{33}} & -\omega + \frac{k_2^2}{\omega \varepsilon_{33}}  & -\frac{k_1 k_2}{\omega \varepsilon_{33}}\\
		-\frac{k_1 k_2}{\omega} - \frac{\omega e_{12}}{\varepsilon_{33}} & \frac{k_1^2}{\omega} - \frac{\omega e_{11}}{\varepsilon_{33}} & -k_2 \frac{\varepsilon_{23}}{\varepsilon_{33}} & k_1 \frac{\varepsilon_{23}}{\varepsilon_{33}}\\
		-\frac{k_2^2}{\omega} + \frac{\omega e_{22}}{\varepsilon_{33}}  & \frac{k_1 k_2}{\omega} + \frac{\omega e_{21}}{\varepsilon_{33}}  & k_2 \frac{\varepsilon_{13}}{\varepsilon_{33}} & -k_1 \frac{\varepsilon_{13}}{\varepsilon_{33}}
	\end{bmatrix},\ 
	\bm{v} = \begin{bmatrix}
	\mathcal{E}_x \\ 
	\mathcal{E}_y \\
	\mathcal{H}_x \\
	\mathcal{H}_y
	\end{bmatrix} =
	\begin{bmatrix}
	\eb\\ \hb
	\end{bmatrix}.
\end{equation}
The spectrum of the problem, $k_3 = k_3 (\bm{k},\omega)$, is solution of the solvability condition of \eqref{eq:M}, which is a 4th degree equation in $k_3$: $\det (\mathbf{M} - k_3 \mathbf{I}) = 0$.  This equation has 4 solutions, $\kappa_n$. For vacuum case, $\varepsilon_{\alpha\beta} = \delta_{\alpha\beta}$, we obtain two double-degenerate roots  $\kappa_{1,2} = +k_z$ and $\kappa_{3,4} = -k_z$. Corresponding eigenvectors read
\begin{equation}\label{eq:vv}
\bm{v}_1^0 = \left[1,0,\frac{k_1 k_2}{-\omega k_z},\frac{k_1^2 + k_z^2}{\omega k_z}\right]^{T}\hspace{-1ex}, \bm{v}_2^0 = \left[0, 1, \frac{k_2^2 + k_z^2}{-\omega k_z}, \frac{k_1 k_2}{\omega k_z}\right]^{T}\hspace{-1ex}, \bm{v}_{3,4}^0 = \left.\bm{v}_{1,2}^0 \right|_{k_z\to -k_z}.
\end{equation}
Then, the general form of the field in vacuum case is a linear combination of these solutions $ \vb^0 = e^{\ib k_z z} \left(v_1^0 \vb_1^0 + v_2^0 \vb_2^0 \right)  + e^{-\ib k_z z} \left( v_3^0 \vb_3^0 + v_4^0 \vb_4^0\right)$, with constants $v_i^0$.

In the non-vacuum case the amplitudes read  
\begin{gather}
\bm{v}_1 = 
\begin{bmatrix}
1 \\
\frac{k_2 (k_1 \xi_3 - k_3 \xi_1) + \omega^2 (k_3 e_{12} - k_1 e_{32})}{k_3 (kk\varepsilon) + k_1 (k_1 \xi_3 - k_3 \xi_1) + \omega^2 (k_1 e_{31} - k_3 e_{11})}\\[1ex]
\frac{- k_1 k_2 (kk\varepsilon) + \omega^2 (k_1 k_3 e_{32} + k_1 k_2 e_{33} - k_3^2 e_{12} - k_2 k_3 e_{13})}{\omega(k_3 (kk\varepsilon) + k_1 (k_1 \xi_3 - k_3 \xi_1) + \omega^2 (k_1 e_{31} - k_3 e_{11}))} \\[1ex]
\frac{-k_2^2 (kk\varepsilon) + \omega^2 (k_1 k_2 (e_{12} + e_{21}) + (k_1^2 + k_3^2) e_{22} + k_2 k_3 (e_{32} + e_{23}) + k_2^2 (e_{11} + e_{33})) - \omega^4 \det\bm{\varepsilon}}{\omega(k_3 (kk\varepsilon) + k_1 (k_1 \xi_3 - k_3 \xi_1) + \omega^2 (k_1 e_{31} - k_3 e_{11}))}
\end{bmatrix}_{k_3 \to \kappa_1},\ann [1ex] 
\bm{v}_2 = 
\begin{bmatrix}
\frac{k_1 (k_2 \xi_3 - k_3 \xi_2) + \omega^2 (k_3 e_{21} + k_2 e_{31})}{k_3 (kk\varepsilon) + k_2 (k_2 \xi_3 - k_3 \xi_2) - \omega^2 (k_2 e_{32} + k_3 e_{22})} \\[1ex]
1\\
\frac{k_1^2 (kk\varepsilon) + \omega^2 (-k_1 k_2 (e_{12} + e_{21}) - (k_2^2 + k_3^2) e_{11} + k_1 k_3 (e_{31} + e_{13}) - k_1^2 (e_{22} + e_{33})) + \omega^4 \det \bm{\varepsilon}}{\omega(k_3 (kk\varepsilon) + k_2 (k_2 \xi_3 - k_3 \xi_2) - \omega^2 (k_2 e_{32} + k_3 e_{22}))} \\[1ex]
\frac{k_1 k_2 (kk\varepsilon) + \omega^2 (-k_1 k_3 e_{23} - k_1 k_2 e_{33} + k_3^2 e_{21} + k_2 k_3 e_{31})}{\omega(k_3 (kk\varepsilon) + k_2 (k_2 \xi_3 - k_3 \xi_2) - \omega^2 (k_2 e_{32} + k_3 e_{22}))} \\[1ex]
\end{bmatrix}_{k_3 \to \kappa_2},\label{eq:vnv}
\nonumber\end{gather}
where $\xi_\alpha = k^\beta \varepsilon_{\beta\alpha}$ and $(kk\varepsilon) = k^\alpha k^\beta \varepsilon_{\alpha\beta}$. Also $\vb_3 = \left.\vb_1 \right|_{k_3 \to \kappa_3}$, $\vb_4 = \left.\vb_2 \right|_{k_3 \to \kappa_4}$. General solution is again a linear combination of these 4 solutions $\vb = \sum_{n=1}^4 e^{\ib \kappa_n z} v_n \vb_n$. 

We are ready now to solve \eqref{eq:S} taking into account boundary conditions \eqref{eq:boundary}. With a somewhat cumbersome calculation we obtain the Fresnel reflection matrices
\begin{eqnarray}
\Rb &=&\frac{-1}{\Delta}
\begin{bmatrix}
\begin{vmatrix}
\vb_1^l & \vb_4^l & \vbh_1^r & \vbh_2^r
\end{vmatrix} 
& \hspace{-1ex}
\begin{vmatrix}
\vb_2^l & \vb_4^l & \vbh_1^r & \vbh_2^r
\end{vmatrix} \\[1ex]
\begin{vmatrix}
\vb_3^l & \vb_1^l & \vbh_1^r & \vbh_2^r
\end{vmatrix}
& \hspace{-1ex}
\begin{vmatrix}
\vb_3^l & \vb_2^l & \vbh_1^r & \vbh_2^r
\end{vmatrix}
\end{bmatrix}\hspace{-1ex}, \Rb' = 
\frac{-1}{\Delta}
\begin{bmatrix}
\begin{vmatrix}
\vb_3^l & \vb_4^l & \vbh_3^r & \vbh_2^r
\end{vmatrix}
&  \hspace{-1ex}
\begin{vmatrix}
\vb_3^l & \vb_4^l & \vbh_4^r & \vbh_2^r
\end{vmatrix}\\[1ex]
\begin{vmatrix}
\vb_3^l & \vb_4^l & \vbh_1^r & \vbh_3^r
\end{vmatrix}
& \hspace{-1ex}
\begin{vmatrix}
\vb_3^l & \vb_4^l & \vbh_1^r & \vbh_4^r
\end{vmatrix}
\end{bmatrix}\hspace{-1ex}, \label{eq:Smatrix}
\end{eqnarray}
where $\Delta = |\vb_3^l \ \vb_4^l \ \vbh_1^r \ \vbh_2^r|$, $\vbh_n^r =  \vb_n^0 + \vb_n^\sigma$, $\vb_n^\sigma = (\bm{0}, -4\pi \ib \sigma_2 \bm{\sigma}_s \eb_n^r)^T$, $\vb_n^l = \vb_n$ and $\sigma_2$ is Pauli matrix. $\bm{\sigma}_s$ is the surface conductivity on the interface.

If the matter is on the right and vacuum is on the left,  we have the same formulas for scattering matrix \eqref{eq:Smatrix}, where for $\vb^l$ we have to use vacuum case \eqref{eq:vv}, and for $\vb^r$ -- expressions for matter \eqref{eq:vnv}. With these changes the conductivity appears within vacuum vectors, only. Therefore, the  Fresnel reflection matrices read
\begin{eqnarray}
\Rb &=& 
\frac{-1}{\Delta'}
\begin{bmatrix}
\begin{vmatrix}
\vbh_1^l & \vbh_4^l & \vb_1^r & \vb_2^r
\end{vmatrix} 
& \hspace{-1ex}
\begin{vmatrix}
\vbh_2^l & \vbh_4^l & \vb_1^r & \vb_2^r
\end{vmatrix} \\[1ex]
\begin{vmatrix}
\vbh_3^l & \vbh_1^l & \vb_1^r & \vb_2^r
\end{vmatrix}
& \hspace{-1ex}
\begin{vmatrix}
\vbh_3^l & \vbh_2^l & \vb_1^r & \vb_2^r
\end{vmatrix}
\end{bmatrix}\hspace{-1ex}, \Rb' = \hspace{-.5ex}
\frac{-1}{\Delta'}
\begin{bmatrix}
\begin{vmatrix}
\vbh_3^l & \vbh_4^l & \vb_3^r & \vb_2^r
\end{vmatrix}
& \hspace{-1ex}
\begin{vmatrix}
\vbh_3^l & \vbh_4^l & \vb_4^r & \vb_2^r
\end{vmatrix}\\[1ex]
\begin{vmatrix}
\vbh_3^l & \vbh_4^l & \vb_1^r & \vb_3^r
\end{vmatrix}
& \hspace{-1ex}
\begin{vmatrix}
\vbh_3^l & \vbh_4^l & \vb_1^r & \vb_4^r
\end{vmatrix}
\end{bmatrix}\hspace{-1ex}, \label{eq:SmatrixN}
\end{eqnarray}
where $\Delta' = |\vbh_3^l \ \vbh_4^l \ \vb_1^r \ \vb_2^r|$, $\vbh_n^l = \vb_n^0 + \vb_n^\sigma$ and $\vb_n^r = \vb_n$. Also, in general, we have to change $\bm{\sigma} \to \bm{\sigma}'$ and $\bm{\varepsilon} \to \bm{\varepsilon}'$.


\section{Conclusion}
We considered the scattering problem for $3D$ extended system (see Fig. \ref{fig:f1}). The system consists of two parts characterized by surface conductivities 
$\bm{\sigma}$, $\bm{\sigma}'$ 
and the bulk dielectric permittivities $ \bm{\varepsilon}$, $\bm{\varepsilon}'$. The Casimir energy for this system has the form given by Eq. \eqref{eq:R}, where the Fresnel matrices are given by Eqs. \eqref{eq:Smatrix} and  \eqref{eq:SmatrixN}.  The limiting cases of uniaxial materials can be shown to coincide with known results\cite{Barash1973,Rosa2008,Antezza2017}.


\section*{Acknowledgments}
The work of N. K. was supported in parts by the grants 2016/03319-6, 2019/10719-9 and 2019/06033-4 of the S\~ao Paulo Research Foundation (FAPESP),  by the RFBR projects 19-02-00496-a.

\end{document}